\documentclass[10pt]{IEEEtran15}
\input{psfig.sty}
\input{epsf.sty}
\usepackage{fancybox}   % For the overlay box
\usepackage{psfig}      % To iclude ".eps" figures

\usepackage{graphicx}
\usepackage{amsmath}
\usepackage{color}
\usepackage{epsfig}
\usepackage{amssymb}

\newcommand{\Define}{\stackrel{\triangle}{=}}

\begin{document}
\twocolumn

\title{\LARGE A Hybrid RTS-BP Algorithm for Improved Detection of Large-MIMO 
$M$-QAM Signals}
\author{Tanumay Datta, N. Srinidhi, A. Chockalingam, and B. Sundar Rajan \\
{\normalsize Department of ECE, Indian Institute of Science,
Bangalore 560012, INDIA}
\vspace{-5mm}
}
\maketitle
\thispagestyle{empty}
\begin{abstract}
Low-complexity near-optimal detection of large-MI\-MO signals has 
attracted recent research. Recently, we proposed a local neighborhood 
search algorithm, namely {\em reactive tabu search} (RTS) algorithm, 
as well as a factor-graph based {\em belief propagation} (BP) algorithm 
for low-complexity large-MIMO detection. The motivation for the present 
work arises from the following two observations on the above two 
algorithms: $i)$ RTS works for general $M$-QAM. Although RTS was shown 
to achieve close to optimal performance for 4-QAM in large dimensions, 
significant performance improvement was still possible for high\-er-order 
QAM (e.g., 16- and 64-QAM). $ii)$ BP also was shown to achieve near-optimal 
performance for large dimensions, but only for $\{\pm 1\}$ alphabet. In 
this paper, we improve the large-MIMO detection performance of higher-order 
QAM signals by using a hybrid algorithm that employs RTS and BP. In 
particular, motivated by the observation that when a detection error 
occurs at the RTS output, the least significant bits (LSB) of the symbols 
are mostly in error, we propose to first reconstruct and cancel the 
interference due to bits other than LSBs at the RTS output and feed 
the interference cancelled received signal to the BP algorithm to improve 
the reliability of the LSBs. The output of the BP is then fed back to 
RTS for the next iteration. Our simulation results show that in a 
$32\times 32$ V-BLAST system, the proposed RTS-BP algorithm performs 
better than RTS by about 3.5 dB at $10^{-3}$ uncoded BER and by about 
2.5 dB at $3\times 10^{-4}$ rate-3/4 turbo coded BER with 64-QAM at the 
same order of complexity as RTS. We also illustrate the performance of 
large-MIMO detection in frequency-selective fading channels.
\end{abstract}
\vspace{-2mm}

\vspace{-4.0mm}
\section{Introduction}
\label{sec1}
\vspace{-2.0mm}
Multiple-input multiple-output (MIMO) systems with large number (e.g., 
tens) of transmit and receive antennas, referred to as `large-MIMO 
systems,' are of interest because of the high capacities/spectral 
efficiencies theoretically predicted in these systems 
\cite{tela99},\cite{paulraj}. Research in low-complexity  receive 
processing (e.g., MIMO detection) techniques that can lead to 
practical realization of large-MIMO systems is both nascent as well 
as promising. For e.g., NTT DoCoMo has already field demonstrated 
a $12\times 12$ V-BLAST system operating at 5 Gbps data rate and 
50 bps/Hz spectral efficiency in 5 GHz band at a mobile speed of 
10 Km/hr \cite{docomo}. Evolution of WiFi standards (evolution from 
IEEE 802.11n to IEEE 802.11ac to achieve multi-gigabit rate 
transmissions in 5 GHz band) now considers $16\times 16$ MIMO operation; 
see $16\times 16$ MIMO indoor channel sounding measurements at 5 GHz 
reported in \cite{dot11ac} for consideration in WiFi standards. Also,
$64 \times 64$ MIMO channel sounding measurements at 5 GHz in indoor 
environments have been reported in \cite{hut}. We note that, while 
RF/antenna technologies/measurements for large-MIMO systems are getting 
matured, there is an increasing need to focus on low-complexity 
algorithms for detection in large-MIMO systems to reap their high 
spectral efficiency benefits. 

In the above context, in our recent works, we have shown that certain 
algorithms from machine learning/artificial intelligence achieve 
near-optimal performance in large-MIMO systems at low complexities
\cite{jsac}-\cite{itw10}\footnote{Similar algorithms have been reported
earlier in the context of multiuser detection in large CDMA systems.}. 
In \cite{jsac}-\cite{stbc}, a local neighborhood search based algorithm,
namely, a {\em likelihood ascent search} (LAS) algorithm, was proposed 
and shown to achieve close to maximum-likelihood (ML) performance in
MIMO systems with several tens of antennas (e.g., $32\times 32$ and
$64\times 64$ MIMO). Subsequently, in \cite{rts},\cite{ldrts}, another 
local search algorithm, namely, {\em reactive tabu search} (RTS) algorithm, 
which performed better than the LAS algorithm through the use of a local 
minima exit strategy was presented\footnote{In \cite{stbc},\cite{ldrts}, 
we compared the performance and complexities of LAS and RTS algorithms
with those of the sphere decoding (SD) variants in \cite{fsd} and
\cite{rsd}, and showed that these SD variants do not scale well for the 
large dimensions considered.}. In \cite{gibbs}, near-ML performance in 
a $50\times 50$ MIMO system was demonstrated using a {\em Gibbs sampling} 
based detection algorithm, where the symbols take values from $\{\pm 1\}$. 
More recently, we, in \cite{itw10}, proposed a factor graph based {\em 
belief propagation} (BP) algorithm for large-MIMO detection, where we 
adopted a Gaussian approximation of the interference (GAI).

The motivation for the present work arises from the following two 
observations on the RTS and BP algorithms in \cite{rts},\cite{ldrts}
and \cite{itw10}: $i)$ RTS works for general $M$-QAM. Although RTS was 
shown to achieve close to ML performance for 4-QAM in large dimensions,
significant performance improvement was still possible for higher-order 
QAM (e.g., 16- and 64-QAM). $ii)$ BP also was shown to achieve 
near-optimal performance for large dimensions, but only for $\{\pm 1\}$ 
alphabet. In this paper, we improve the large-MIMO detection performance 
of higher-order QAM signals by using a hybrid algorithm that employs RTS 
and BP. In particular, we observed that when a detection error occurs at 
the RTS output, the least significant bits (LSB) of the symbols are mostly 
in error. Motivated by this observation, we propose to first reconstruct 
and cancel the interference due to bits other than the LSBs at the RTS 
output and feed the interference cancelled received signal to the BP 
algorithm to improve the reliability of the LSBs. The output of the BP 
is then fed back to the RTS for the next iteration. Our simulation 
results show that the proposed RTS-BP algorithm achieves better uncoded 
as well as coded BER performance compared to those achieved by RTS in 
large-MIMO systems with higher-order QAM (e.g., RTS-BP performs better
by about 3.5 dB at $10^{-3}$ uncoded BER and by about 2.5 dB at 
$3\times 10^{-4}$ rate-3/4 turbo coded BER in $32\times 32$ V-BLAST 
with 64-QAM) at the same order of complexity as RTS.

The rest of this paper is organized as follows. In Sec. \ref{sec2}, we 
introduce the RTS and BP algorithms in \cite{rts},\cite{ldrts} and 
\cite{itw10} and the motivation for the current work. The proposed 
hybrid RTS-BP algorithm and its performance are presented in Secs. 
\ref{sec3} and \ref{sec4}. Conclusions are given in Sec. \ref{sec5}.

\vspace{-3.0mm}
\section{RTS and BP Algorithms for Large-MIMO Detection}
\label{sec2}
\vspace{-1.0mm}
Consider a $N_t\times N_r$ V-BLAST MIMO system whose received signal 
vector, ${\bf y}_c \in {\mathbb C}^{N_r}$, is of the form 
\vspace{-2mm}
\begin{eqnarray}
{\bf y}_c & = & {\bf H}_c{\bf x}_c + {\bf n}_c,
\label{eqn1}
\end{eqnarray}

\vspace{-4mm}
where ${\bf x}_c \in {\mathbb C}^{N_t}$ is the symbol vector
transmitted, ${\bf H}_c \in {\mathbb C}^{N_r \times N_t}$ is the
channel gain matrix, and ${\bf n}_c \in {\mathbb C}^{N_r}$ is 
the noise vector whose entries are modeled as i.i.d 
${\mathbb C}{\mathcal N}(0,\sigma^2)$. Assuming rich scattering, 
we model the entries of ${\bf H}_c$ as i.i.d $\mathcal C \mathcal N(0,1)$. 
Each element of ${\bf x}_c$ is an $M$-PAM or $M$-QAM symbol. $M$-PAM 
symbols take values from $\{A_m, m=1,2,\\\cdots,M\}$, where 
$A_m=(2m-1-M)$, and $M$-QAM is nothing but two PAMs in quadrature. 
As in \cite{isit08}, we convert (\ref{eqn1}) into a real-valued 
system model, given by
\vspace{-2mm}
\begin{eqnarray}
{\bf y} & = & {\bf H}{\bf x} + {\bf n},
\label{eqn2}
\end{eqnarray}

\vspace{-4mm}
where ${\bf H} \in {\mathbb R}^{2N_r \times 2N_t}$,
${\bf y} \in {\mathbb R}^{2N_r}$,
${\bf x} \in {\mathbb R}^{2N_t}$, 
${\bf n} \in {\mathbb R}^{2N_r}$.
For $M$-QAM, {\small $[x_1,\cdots,x_{N_t}]$} can viewed to be from an
underlying $M$-PAM signal set, and so is $[x_{N_t+1},\cdots,x_{2N_t}]$. Let
$\mathbb A_i$ denote the $M$-PAM signal set from which $x_i$ takes values,
$i=1,2,\cdots,2N_t$. 
Defining a $2N_t$-dimensional signal space $\mathbb S$ to be the
Cartesian product of $\mathbb A_1$ to $\mathbb A_{2N_t}$, the
ML solution vector, ${\bf x}_{ML}$, is given by
\vspace{-2mm}
\begin{eqnarray}
\label{MLdetection}
{\bf x}_{ML} & = & {\mbox{arg min}\atop{{\bf x} \in {\mathbb S}}}
\hspace{2mm} \Vert {\bf y} - {\bf H}{\bf x} \Vert ^2,
\end{eqnarray}
whose complexity is exponential in $N_t$. The RTS algorithm in 
\cite{rts},\cite{ldrts} is a low-complexity algorithm, which minimizes
the ML metric in (\ref{MLdetection}) through a local neighborhood 
search.

\vspace{-4mm}
\subsection{RTS Algorithm} 
\label{rts_sec}
\vspace{-1mm}
A detailed description of the RTS algorithm for large-MIMO detection
is available in \cite{rts},\cite{ldrts}. Here, we present a brief 
summary of the key aspects of the algorithm, and its 16- and 64-QAM
performance that motivates the current work.

The RTS algorithm starts with an initial solution vector, defines
a neighborhood around it (i.e., defines a set of neighboring vectors
based on a neighborhood criteria),
and moves to the best vector among the neighboring vectors (even if the
best neighboring vector is worse, in terms of likelihood, than the
current solution vector; this allows the algorithm to escape from local
minima). This process is continued for a certain number of iterations,
after which the algorithm is terminated and the best among the
solution vectors in all the iterations is declared as the final solution
vector. In defining the neighborhood of the solution vector in a given
iteration, the algorithm attempts to avoid cycling by making the moves to
solution vectors of the past few iterations as `tabu' (i.e., prohibits these
moves), which ensures efficient search of the solution space. The number
of these past iterations is parametrized as the `tabu period.' The search
is referred to as fixed tabu search if the tabu period is kept constant.
If the tabu period is dynamically changed (e.g., increase the tabu period
if more repetitions of the solution vectors are observed in the search
path), then the search is called reactive tabu search. We consider reactive
tabu search because of its robustness (choice of a good fixed tabu period 
can be tedious). The per-symbol complexity of RTS for detection of V-BLAST 
signals is $O(N_tN_r)$. 

\subsubsection{Motivation of Current Work}
\begin{figure}
\epsfysize=7.2cm
\epsfxsize=9.60cm
\hspace{-6mm}
\epsfbox{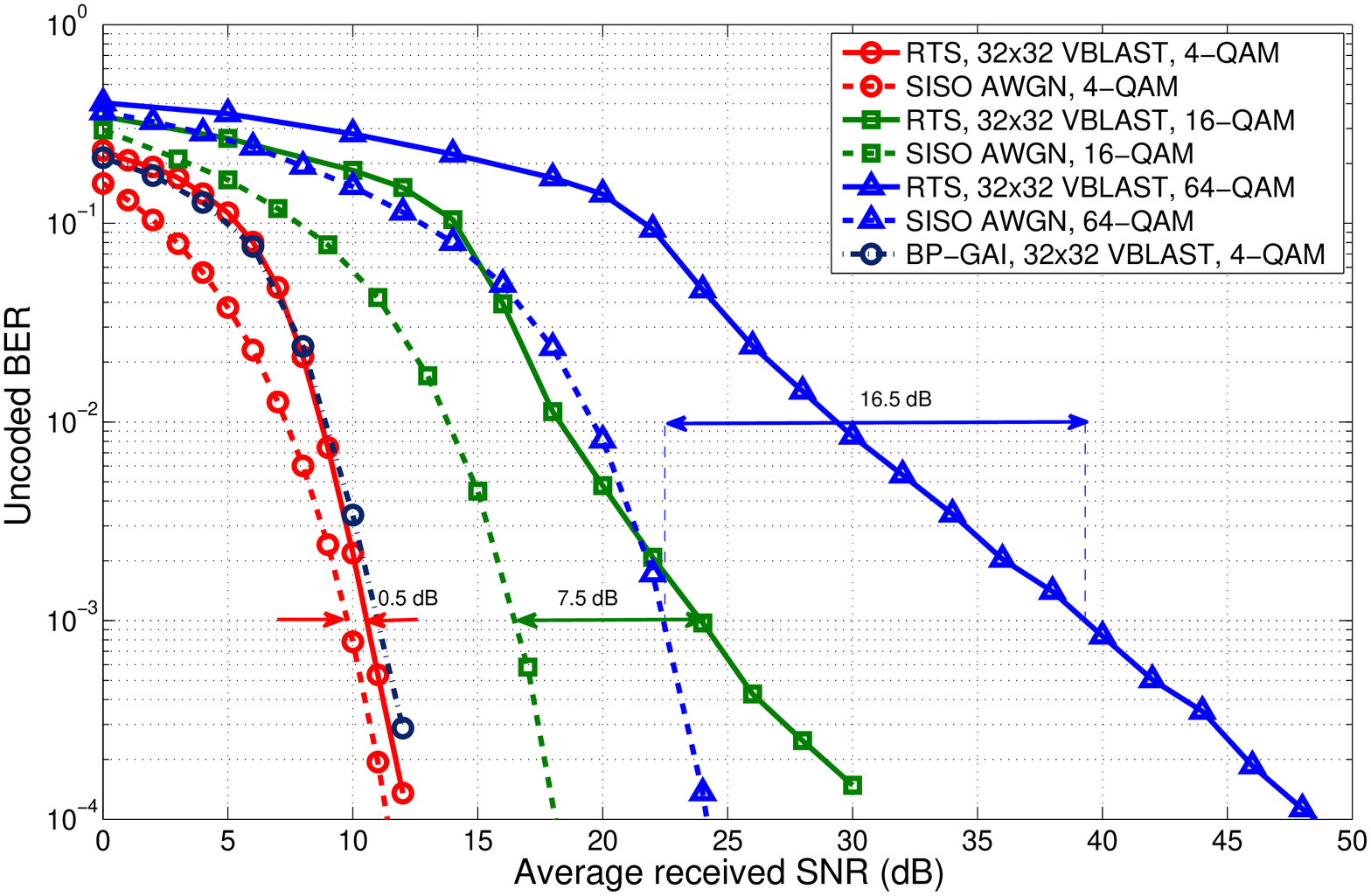}
\vspace{-9.5mm}
\caption{Uncoded BER performance of RTS algorithm in $32\times 32$
V-BLAST for 4-, 16-, 64-QAM. 
{\em Performance improvement is possible in 16-, 64-QAM}.
}
\vspace{-4.5mm}
\label{fig1}
\end{figure}

Figure \ref{fig1} shows the uncoded BER performance of RTS using the
algorithm parameters optimized through simulations for 4-, 16-, and 
64-QAM in a $32\times 32$ V-BLAST system. As lower bounds on the error 
performance in MIMO, the SISO AWGN performance for 4-, 16-, and 64-QAM 
are also plotted. It can be seen that, in the case of 4-QAM, the RTS 
performance is just about 0.5 dB away from the SISO AWGN performance 
at $10^{-3}$ BER. However, the gap between RTS performance and SISO 
AWGN performance at $10^{-3}$ BER widens for 16-QAM and 64-QAM; the 
gap is 7.5 dB for 16-QAM and 16.5 dB for 64-QAM. This gap can be viewed
as a potential indicator of the amount of improvement in performance
possible further. A more appropriate indicator will be the gap between RTS 
performance and the ML performance. Since simulation of sphere decoding
(SD) of $32\times 32$ V-BLAST with 16- and 64-QAM (64 real dimensions) 
is computationally intensive, we do not show the SD (ML) performance. 
Nevertheless, the widening gap of RTS Performance from SISO AWGN 
performance for 16- and 64-QAM seen in Fig. \ref{fig1} motivated us 
to explore improved algorithms to achieve better performance than RTS 
performance for higher-order QAM. 

\vspace{-3.5mm}
\subsection{BP Algorithm Based on GAI}
\label{sec_bp}
\vspace{-1.5mm}
In \cite{itw10}, we presented a detection algorithm based on BP on 
factor graphs of MIMO systems. In (\ref{eqn2}), each entry of the 
vector ${\bf y}$ is treated as a function node (observation
node), and each symbol, $x_i \in \{\pm 1\}$, as a variable node. 
A key ingredient in 
the BP algorithm in \cite{itw10}, which contributes to its low 
complexity, is the Gaussian approximation of interference (GAI), 
where the interference plus noise term, $z_{ik}$, in

\begin{figure}
\begin{minipage}{2.4cm}
\begin{center}
\epsfysize=3.5cm
\epsfxsize=4.2cm
\epsfbox{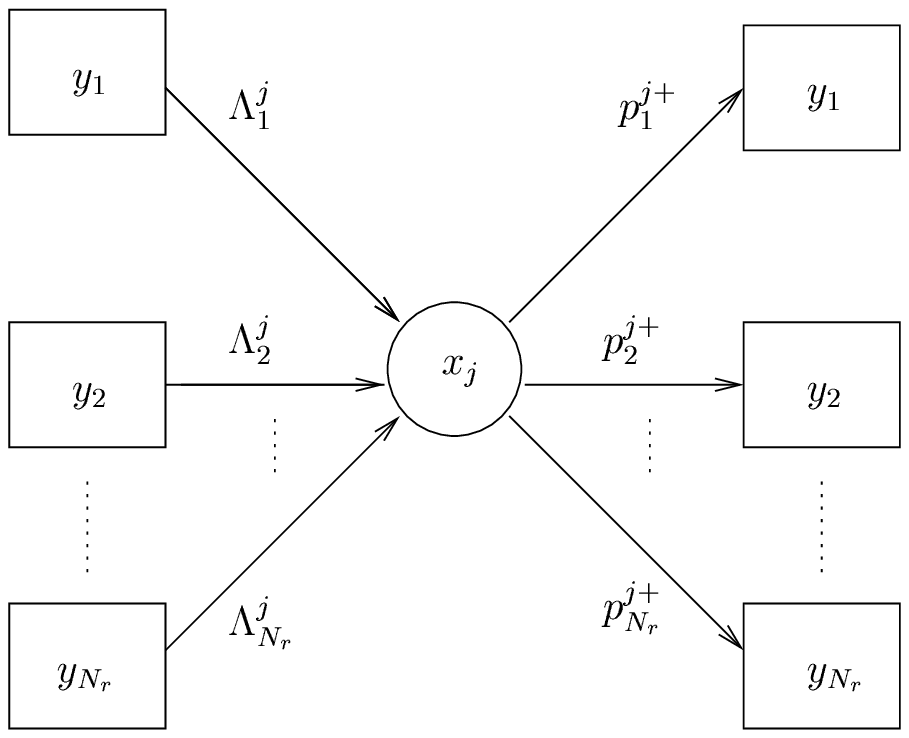}
\end{center}
\end{minipage}\hspace{20mm}
\begin{minipage}{2.4cm}
\begin{center}
\epsfysize=3.5cm
\epsfxsize=4.2cm
\epsfbox{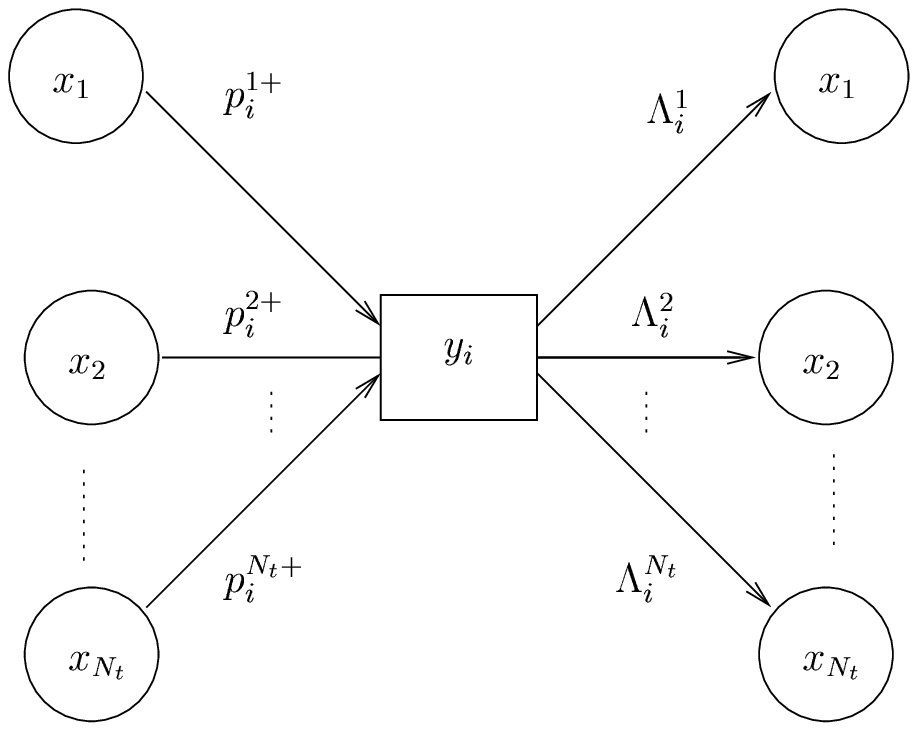}
\end{center}
\end{minipage}
\vspace{3mm}
\caption{Message passing between variable nodes and observation nodes.}
\label{fig2}
\vspace{-4mm}
\end{figure}

\vspace{-6mm}
{\small
\begin{eqnarray}
y_i & = & h_{ik}x_k + \underbrace{\overbrace{\sum_{j=1,j\ne k}^{2N_t}h_{ij} x_j}^{interference} + \,\,n_i}_{\Define \,\, z_{ik} \, 
}, 
\label{model2}
\end{eqnarray}
}

is modeled as
{\small $\mathbb{C}{\cal N}(\mu_{z_{ik}},\sigma^2_{z_{ik}})$} with 
{\small $\mu_{z_{ik}} =  \sum_{j=1,j\ne k}^{N_t}h_{ij}\mathbb{E}(x_j)$}, 

and
{\small $\sigma^2_{z_{ik}} = \sum_{j=1,j\ne k}^{2N_t}|h_{ij}|^2 \, \mbox{Var}(x_j) + \frac{\sigma^2}{2}$}, 
where $h_{ij}$ is the $(i,j)$th element in ${\bf H}$. With 
$x_i$'s $\in \{\pm 1\}$, the log-likelihood ratio (LLR) of $x_k$ 
at observation node $i$, denoted by $\Lambda_{i}^{k}$, is 
{\small
\begin{eqnarray}
\hspace{1mm}
\Lambda_{i}^{k} \,\, = \,\, \log\frac{p(y_i|{\bf H},x_k=1)}{p(y_i|{\bf H},x_k=-1)} \,\, = \,\, \frac{2}{\sigma_{z_{ik}}^2} \Re\left(h_{ik}^*(y_i-\mu_{z_{ik}})\right).
\label{LLRform}
\label{LLR} 
\end{eqnarray}
}
The LLR values computed at the observation nodes are passed to
the variable nodes (as shown in Fig. \ref{fig2}). Using these LLRs, 
the variable nodes 
compute the probabilities
\vspace{-2mm}
\begin{eqnarray}
\hspace{-6mm}
p_i^{k+} & \Define & p_i(x_k=+1|{\bf y}) \,\, = \,\, \frac{\mbox{exp}(\sum_{l\neq i}\Lambda_l^k)}{1 + \mbox{exp}(\sum_{l\neq i}\Lambda_l^k)}, 
\label{prob}
\end{eqnarray}
and pass them back to the observation nodes (Fig. \ref{fig2}). This message 
passing is carried out for a certain number of iterations. 
At the end, $x_k$ is detected as 
\vspace{-2mm}
\begin{eqnarray}
\widehat{x}_k & = & \mbox{sgn}\Big(\sum_{i=1}^{2N_r}\Lambda_i^k \Big).
\end{eqnarray}
It has been shown in \cite{itw10} that this BP algorithm with GAI, like 
LAS and RTS algorithms, exhibits `large-system behavior,' where the bit
error performance improves with increasing number of dimensions.
In Fig. \ref{fig1}, the uncoded BER performance of this BP algorithm 
for 4-QAM \big(input data vector of size $2N_t$ with elements from 
$\{\pm 1\}$\big) in $32\times 32$ V-BLAST is also plotted. We can see
that the performance is almost the same as that of RTS. In terms of
complexity, the BP algorithm has the advantage of no need to compute 
an initial solution vector and ${\bf H}^T{\bf H}$, which is required 
in RTS. The per-symbol complexity of the BP algorithm for detection in
V-BLAST is $O(N_t)$. A limitation with this BP approach is that it is 
not for general $M$-QAM. However, its good performance with $\{\pm 1\}$
alphabet at lower complexities than RTS can be exploited to improve the 
higher-order QAM performance of RTS, as proposed in the following section.

\vspace{-3.0mm}
\section{Proposed Hybrid RTS-BP Algorithm for Large-MIMO Detection}
\vspace{-1.0mm}
\label{sec3}
In this section, we highlight the rationale behind the hybrid
RTS-BP approach and present the proposed algorithm.  

{\em Why Hybrid RTS-BP?}

The proposed hybrid RTS-BP approach is motivated by the the following 
observation we made in our RTS BER simulations. We observed that, at 
moderate to high SNRs, when an RTS output vector is in error, the 
least significant bits (LSB) of the data symbols are more likely to 
be in error than other bits. An analytical reasoning for this behavior 
can be given as follows.

Let ${\bf x}$ be the transmit vector and $\widehat{{\bf x}}$ be the 
corresponding output of the RTS detector. Let 
${\mathbb A} = \{a_1,a_2,\cdots,a_M\}$ denote the $M$-PAM alphabet 
that $x_i$'s take values from. Consider the symbol-to-bit mapping, 
where we can write the value of each entry of $\widehat{{\bf x}}$ as 
a linear combination of its constituent bits as
\begin{eqnarray}
\label{linearComb}
\widehat{x}_i&=&\sum_{j=0}^{N-1} 2^j \, \widehat{b}_i^{(j)}, \,\,\,\,\,\, i=1,\cdots,2N_t,
\end{eqnarray}
where $N=\log_2 M$ and $\widehat{b}_i^{(j)} \in \{\pm 1\}$. We note that 
the RTS algorithm outputs a local minima as the solution vector. So, 
$\widehat{\bf{x}}$, being a local minima, satisfies the following conditions: 
\begin{eqnarray}
\Vert {\bf y}-{\bf H}\widehat{\bf x}\Vert^{2} \, \le \, \Vert{\bf y}-{\bf H}(\widehat{\bf x}+\lambda_i\textbf{e}_i)\Vert^{2}, \quad \forall i=1,\cdots 2N_t, 
\label{localminima}
\end{eqnarray}
where $\lambda_i = (a_q - \widehat{x}_i), q=1,\cdots,M$, 
and $\textbf{e}_i$ denotes the $i$th column of the identity matrix. 
Defining ${\bf F} \Define {\bf H}^T{\bf H}$, 
${\bf r} \Define {\bf H}\widehat{{\bf x}}$, and denoting the $i$th
column of ${\bf H}$ as ${\bf h}_i$,  the conditions in 
(\ref{localminima}) reduce to
\vspace{-3mm}
\begin{eqnarray}
2\lambda_i{\bf y}^{T}{\bf h}_i & \le & 2\lambda_i{\bf r}^T {\bf h}_i + \lambda_i^2 f_{ii},
\label{reducedcondition}
\end{eqnarray}
where $f_{ij}$ denotes the $(i,j)$th element of ${\bf F}$.
Under moderate to high SNR conditions, ignoring the noise, 
(\ref{reducedcondition}) can be further reduced to
\vspace{-1mm}
\begin{eqnarray}
2({\bf x}-\widehat{\bf x})^T {\bf f}_i \, \mbox{sgn}(\lambda_i) & \le & \lambda_i f_{ii} \, \mbox{sgn}(\lambda_i),
\label{finalcondn1}
\end{eqnarray}
where ${\bf f}_i$ denotes the $i$th column of {\bf F}.
For Rayleigh fading, $f_{ii}$ is chi-square distributed with $2N_t$
degrees of freedom with mean $N_t$. Approximating the distribution of 
$f_{ij}$ to be normal with mean zero and variance $\frac{N_t}{4}$ for
$i\ne j$
by central limit theorem, we can drop the $\mbox{sgn}(\lambda_i)$ in 
(\ref{finalcondn1}). Using the fact that the minimum value of 
$|\lambda_i|$ is 2, (\ref{finalcondn1}) can be simplified as 
\vspace{-1mm}
\begin{eqnarray}
\sum_{x_j\ne \widehat{x}_{j}} \Delta_j f_{ij} & \le & f_{ii},
\label{finalcondn2}
\end{eqnarray}
where $\Delta_j=x_j-\widehat{x}_j$. 
Also, if $x_i=\widehat{x}_i$, by the normal approximation in the above
\vspace{-2mm}
\begin{eqnarray}
\sum_{x_j\ne \hat{x}_{j}} \Delta_j f_{ij} & \sim &
{\mathcal N}\Big(0,\frac{N_t}{4}{\sum_{{x_j} \ne \hat{x}_{j}}}\Delta_j^2\Big).
\label{finalline}
\end{eqnarray}
Now, the LHS in (\ref{finalcondn2}) being normal with variance proportional
to $\Delta_j^2$ and the RHS being positive, it can be seen that $\Delta_i$, 
$\forall i$ take smaller values with higher probability. Hence, the symbols 
of $\widehat{\bf x}$ are nearest Euclidean neighbors of their corresponding
symbols of the global minima with high probability\footnote{Because 
$x_i$'s and $\widehat{x}_i$'s take values from $M$-PAM alphabet, 
$\widehat{x}_i$ is said to be the Euclidean nearest neighbor of 
$x_i$ if $|x_i-\widehat{x}_i| = 2$.}. 
Now, because of the symbol-to-bit mapping in (\ref{linearComb}), 
$\widehat{x}_i$ will differ from its nearest Euclidean neighbors
certainly in the LSB position, and may or may not differ in other
bit positions. Consequently, 
the LSBs of the symbols in the RTS output $\widehat{{\bf x}}$ are
least reliable.

The above observation then led us to consider improving the reliability 
of the LSBs of the RTS output using the BP algorithm in \cite{itw10}, 
and iterate between RTS and BP as follows. 

\begin{figure}
\epsfysize=2.50cm
\epsfxsize=8.4cm
\hspace{1mm}
\epsfbox{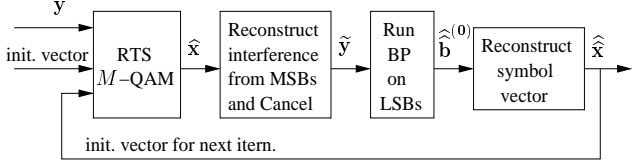}
\vspace{-4mm}
\caption{Proposed hybrid RTS-BP algorithm.}
\vspace{-2mm}
\label{fig3}
\end{figure}

{\em Proposed Hybrid RTS-BP Algorithm:}

Figure \ref{fig3} shows the block schematic of the proposed hybrid
RTS-BP algorithm. The following four steps constitute the proposed 
algorithm.
\begin{itemize}
\item {\em Step 1:}
Obtain $\widehat{{\bf x}}$ using the RTS algorithm. Obtain the
output bits $\widehat{b}_i^{(j)}$, $i=1,\cdots,2N_t$,
$j=0,\cdots,N-1$, from $\widehat{{\bf x}}$ and (\ref{linearComb}).
\item {\em Step 2:} 
Using the $\widehat{b}_i^{(j)}$'s from Step 1, reconstruct the 
interference from all bits other than the LSBs \big(i.e., interference
from all bits other than $\widehat{b}_i^{(0)}$'s\big) as 
\begin{eqnarray}
\widetilde{{\bf I}} & = & \sum_{j=1}^{N-1} 2^{j}\, {\bf H} \, \widehat{{\bf b}}^{(j)},
\label{intf}
\end{eqnarray}
where $\widehat{{\bf b}}^{(j)} = \big[\widehat{b}_1^{(j)}, \widehat{b}_2^{(j)}, \ldots, \widehat{b}_{2N_t}^{(j)} \big]^T$. Cancel the reconstructed
interference in (\ref{intf}) from {\bf y} as
\begin{eqnarray}
\widetilde{{\bf y}} & = & {\bf y} - \widetilde{{\bf I}}.
\end{eqnarray}
\item {\em Step 3:} 
Run the BP-GAI algorithm in Sec. \ref{sec_bp} on the vector
$\widetilde{{\bf y}}$ in Step 2, and obtain an estimate of the 
LSBs. Denote this LSB output vector from BP as 
$\widehat{\widehat{\bf b}}^{(0)}$.
Now, using $\widehat{\widehat{\bf b}}^{(0)}$ from the BP output,
and the $\widehat{\bf b}^{(j)}$, $j=1,\cdots,N-1$ from the RTS
output in Step 1, reconstruct the symbol vector as
\begin{eqnarray}
\widehat{\widehat{{\bf x}}} & = &  \widehat{\widehat{\bf b}}^{(0)} \,+ \sum_{j=1}^{N-1} 2^{j}\, \, \widehat{{\bf b}}^{(j)}.
\end{eqnarray}
\item {\em Step 4:} 
Repeat Steps 1 to 3 using $\widehat{\widehat{{\bf x}}}$
as the initial vector to the RTS algorithm. 
\end{itemize}
The algorithm is stopped after a certain number of iterations between 
RTS and BP. Our simulations showed that two iterations between RTS and 
BP are adequate to achieve good improvement; more than two iterations 
resulted in only marginal improvement for the system parameters 
considered in the simulations. Since the complexity of BP part of
RTS-BP is less than that of the RTS part, the order of complexity of 
RTS-BP is same as that of RTS.

\begin{figure}
\epsfysize=7.3cm
\epsfxsize=9.8cm
\hspace{-6mm}
\epsfbox{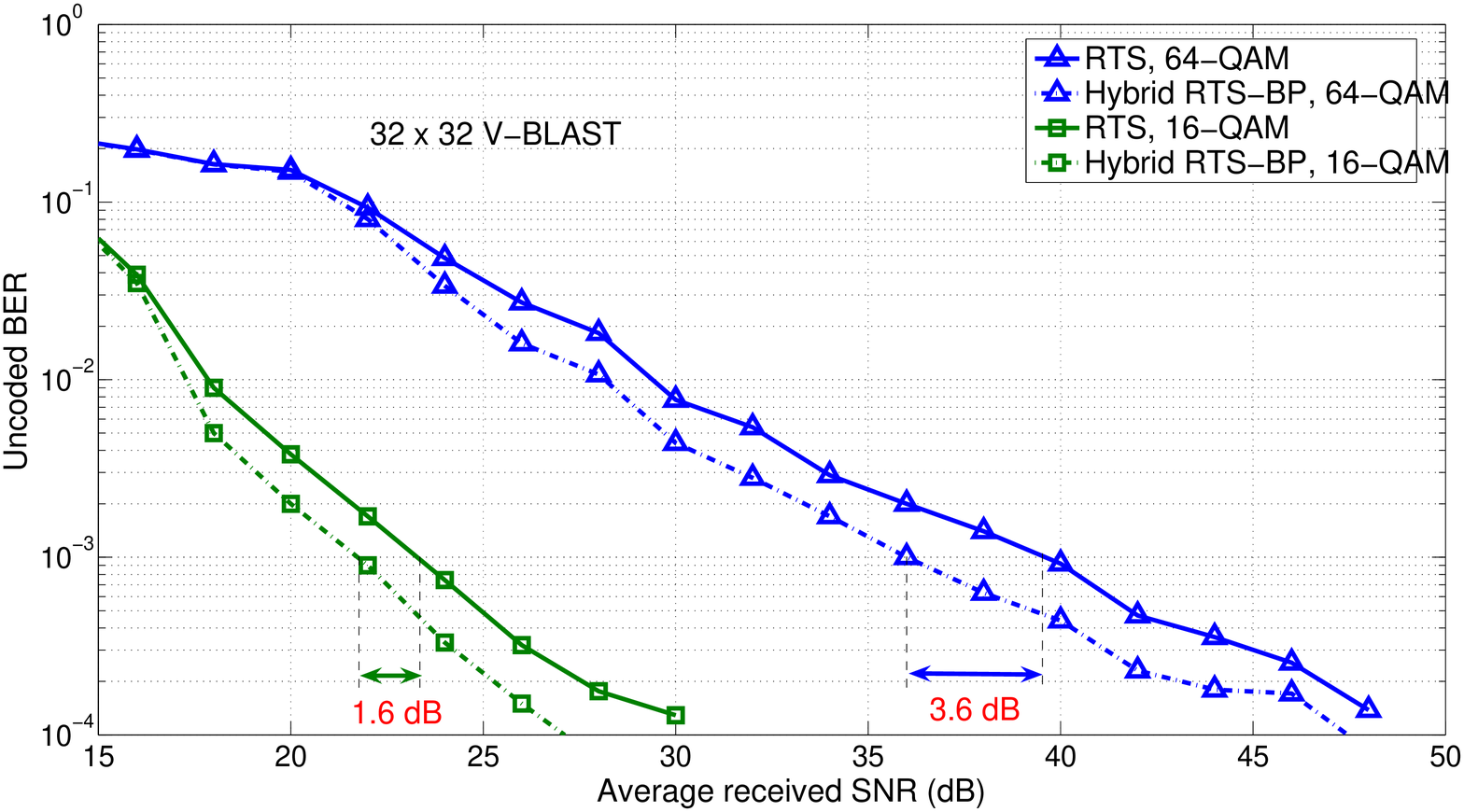}
\vspace{-8mm}
\caption{Uncoded BER comparison between the proposed hybrid RTS-BP and 
the RTS for 16- and 64-QAM in $32\times 32$ V-BLAST. {\em RTS-BP performs 
3.6 dB better than RTS at $10^{-3}$ BER for 64-QAM}.}
\vspace{-2mm}
\label{fig4}
\end{figure}

\vspace{-3.0mm}
\section{BER Performance of the Hybrid RTS-BP Detector }
\vspace{-1.0mm}
\label{sec4}
In this section, we present the uncoded and coded BER performance of
the proposed RTS-BP algorithm evaluated through simulations. Perfect
knowledge of ${\bf H}$ is assumed at the receiver. 

\begin{figure}
\epsfysize=7.3cm
\epsfxsize=9.8cm
\hspace{-6mm}
\epsfbox{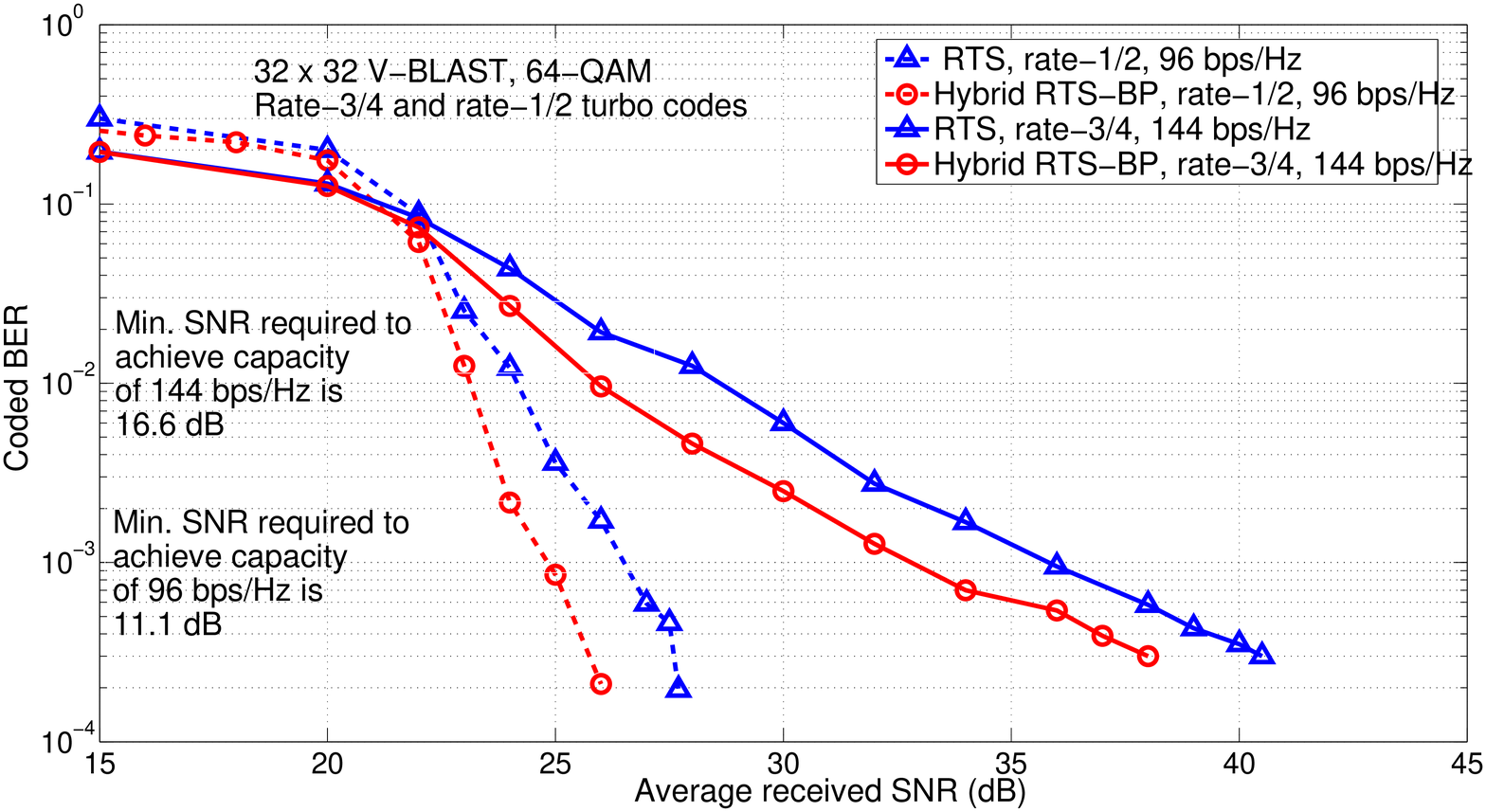}
\vspace{-8mm}
\caption{Coded BER performance comparison between the proposed hybrid 
RTS-BP and the RTS in $32\times 32$ V-BLAST with 64-QAM, $i)$ rate-1/2 
turbo code (96 bps/Hz), $ii)$ rate-3/4 turbo code (144 bps/Hz). {\em RTS-BP
performs better by about 1.5 dB and 2.5 dB, respectively, at these spectral
efficiencies, at $3\times 10^{-4}$ coded BER.}} 
\vspace{-3mm}
\label{fig5}
\end{figure}

{\em Performance in large V-BLAST Systems:}
Figure \ref{fig4} shows the uncoded BER performance of $32\times 32$ 
V-BLAST with 16- and 64-QAM. Performance of both RTS-BP as well as RTS 
are shown. It can be seen that, at an uncoded BER of $10^{-3}$, RTS-BP 
performs better than RTS by about 3.6 dB for 64-QAM and by about 1.6 dB 
for 16-QAM. This illustrates the effectiveness of the proposed hybrid
RTS-BP approach. Also, this improvement in uncoded BER is found to result 
in improved coded BER as well, as illustrated in Fig. \ref{fig5}. In 
Fig. \ref{fig5}, we have plotted the turbo coded BER of RTS-BP and RTS 
in $32\times 32$ V-BLAST with 64-QAM for rate-1/2 (96 bps/Hz) and 
rate-3/4 (144 bps/Hz) turbo codes. It can be seen that, at a coded
BER of $3\times 10^{-4}$, RTS-BP performs better than RTS by about 1.5 dB 
at 96 bps/Hz and by about 2.5 dB at 144 bps/Hz. 

\begin{figure}
\epsfysize=7.3cm
\epsfxsize=10.0cm
\hspace{-8mm}
\epsfbox{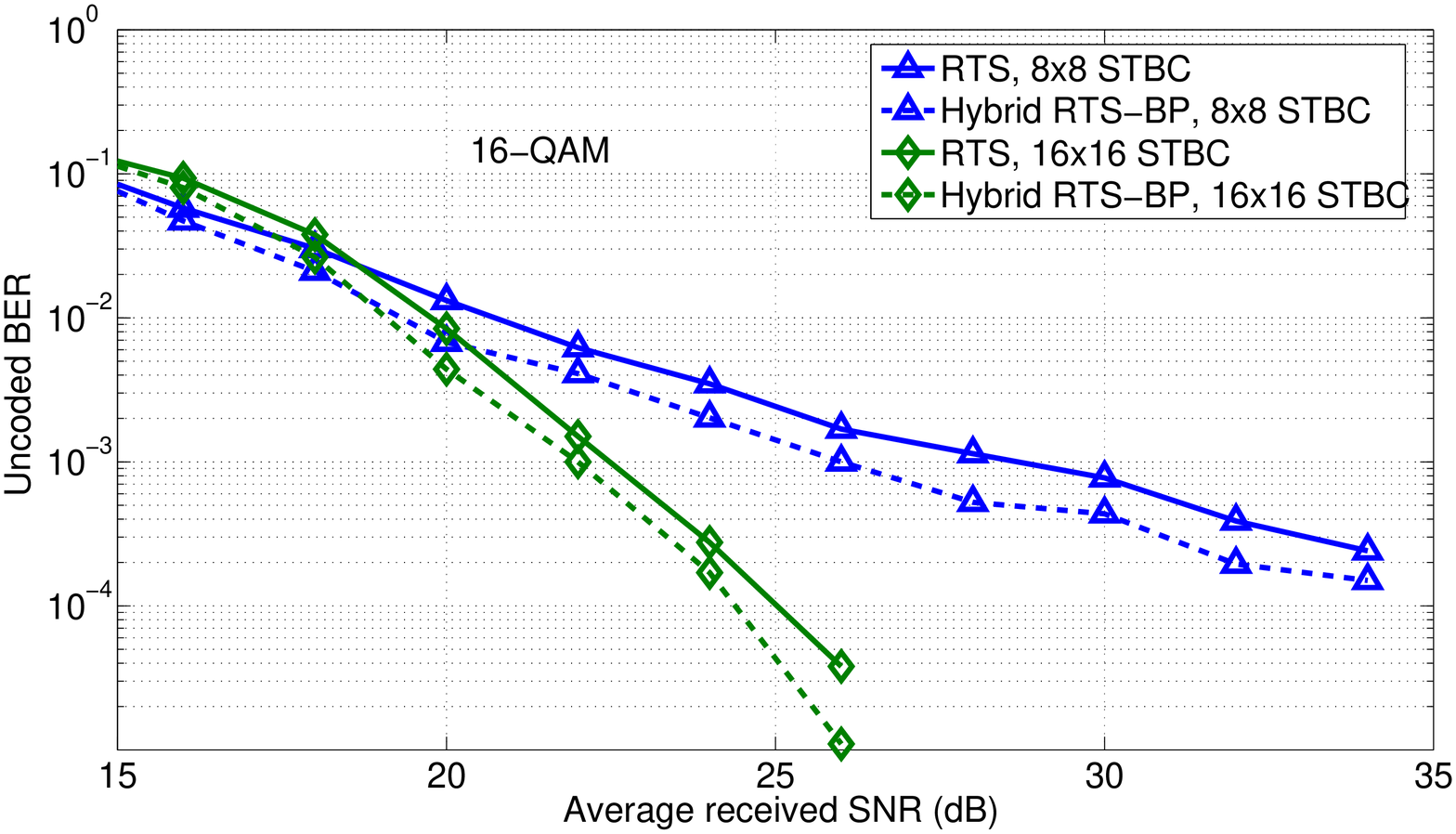}
\vspace{-8mm}
\caption{Uncoded BER performance comparison between the hybrid RTS-BP 
and the RTS for 16-QAM in $8\times 8$ and $16\times 16$ non-orthogonal
STBCs from CDA in [15].}
\vspace{-4mm}
\label{fig6}
\end{figure}

{\em Performance in large non-orthogonal STBC MIMO systems:}
We also evaluated the BER performance of large non-orthogo\-nal STBC MIMO 
systems with higher-order QAM using RTS-BP detection. Figure \ref{fig6} 
shows the uncoded BER of $8\times 8$ and $16\times 16$ non-orthogonal
STBC from cyclic division algebra \cite{bsr} for 16-QAM. Here again,
we can see that RTS-BP achieves better performance than RTS. 

\begin{figure}
\epsfysize=7.3cm
\epsfxsize=10.0cm
\hspace{-8mm}
\epsfbox{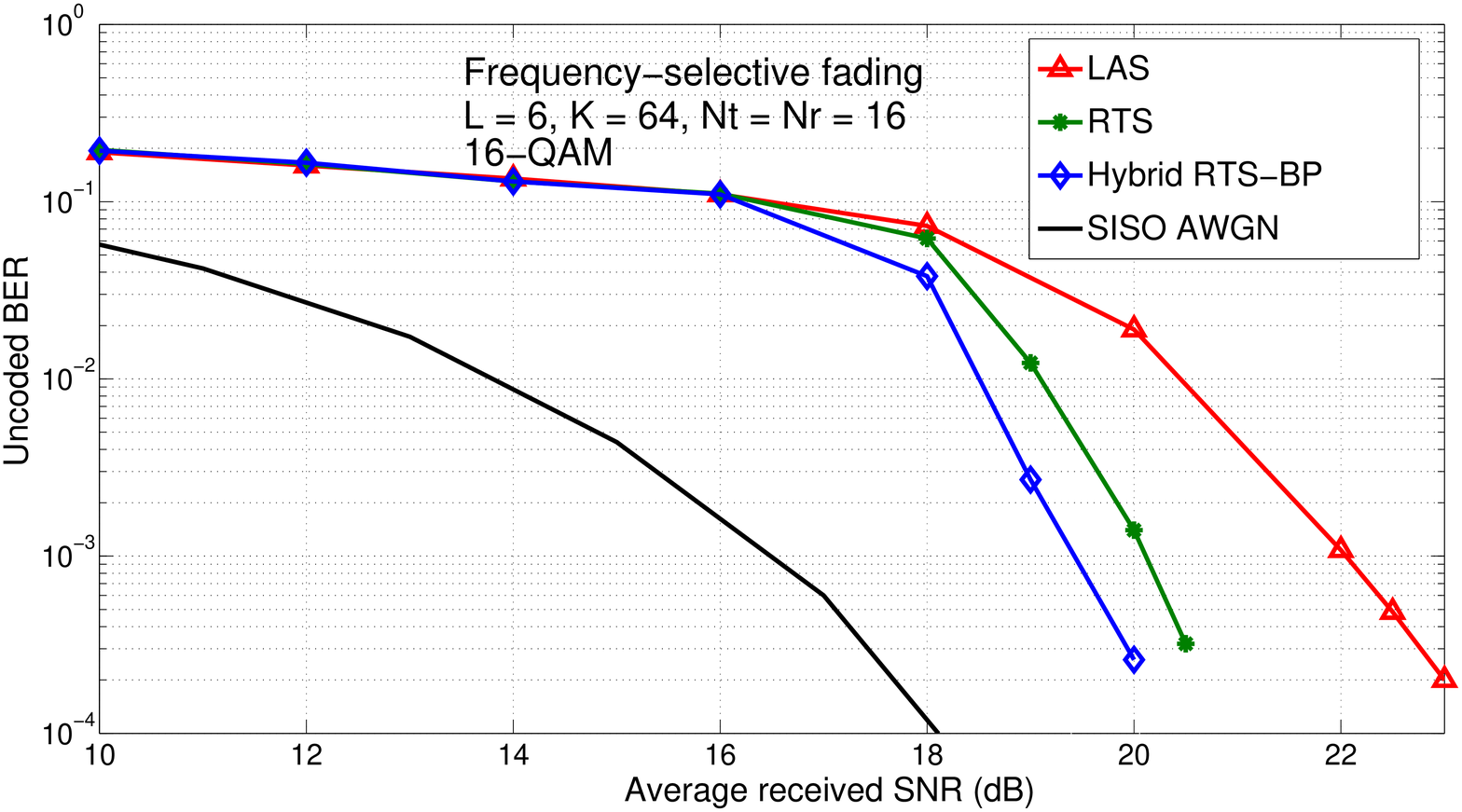}
\vspace{-8mm}
\caption{Uncoded BER performance comparison between the hybrid RTS-BP,
RTS, and LAS algorithms in $16\times 16$ V-BLAST with 16-QAM in 
frequency-selective fading with $L=6$, $K=64$, and uniform power-delay 
profile.}
\vspace{-4mm}
\label{fig7}
\end{figure}

{\em Performance in frequency-selective large V-BLAST systems:}
We note that the performance plots in Figs. \ref{fig4} to \ref{fig6}
are for frequency-flat fading, which could be the fading scenario 
in MIMO-OFDM systems where a frequency-selective fading channel is 
converted to frequency-flat channels on multiple subcarriers. 
RTS-BP, RTS, and LAS algorithms, being suited to work well in large 
dimensions, can be applied to equalize signals in frequency-selective 
channels in large-MIMO systems. Following the equivalent real-valued 
system model of the form in (\ref{eqn2}) for frequency-selective MIMO 
systems developed in \cite{isi_gcom09}, we evaluated the performance of 
RTS-BP, RTS, and LAS algorithms in $16\times 16$ V-BLAST  with 16-QAM
on a frequency 
selective channel with $L=6$ equal energy multipath components 
and $K=64$ symbols per frame. Figure \ref{fig7} shows the superior
performance of the RTS-BP algorithm over the RTS and LAS algorithms in
this frequency-selective $16\times 16$ large-MIMO system with 16-QAM.

\vspace{-3.0mm}
\section{Conclusions}
\vspace{-1.0mm}
\label{sec5}
We proposed a hybrid algorithm that exploited the good features of 
the RTS and BP algorithms to achieve improved bit error performance
and nearness to capacity performance for $M$-QAM signals in large-MIMO
systems at practically affordable low complexities. We illustrated the 
performance gains of the proposed hybrid approach over the RTS algorithm
in flat-fading as well as frequency-selective fading for large V-BLAST 
as well as large non-orthogonal STBC MIMO systems. We note (e.g., from 
the performance plots for 64-QAM in Figs. \ref{fig1} and \ref{fig5}) that 
further improvement in performance beyond what is achieved by the proposed 
hybrid RTS-BP algorithm could be possible. Investigation of alternate 
detection strategies to achieve this possible improvement is a subject 
for further investigation.

\end{document}